\documentclass[conference]{IEEEtran}
\usepackage{amsmath}
\usepackage{graphicx}

\usepackage{amssymb}
\usepackage[margin=0.5in]{geometry}
\usepackage{caption}
\usepackage{subcaption}
\usepackage{bbm}
\usepackage{float}
\usepackage{enumitem}
\IEEEoverridecommandlockouts

\IEEEoverridecommandlockouts
\begin{document}
\title{A New Algorithm for Distributed Nonparametric Sequential Detection\thanks{This work was partially supported by a grant from ANRC.}}
\author{\IEEEauthorblockN{Shouvik Ganguly, K. R. Sahasranand and Vinod Sharma}
\IEEEauthorblockA{Department of Electrical Communication Engineering\\ Indian Institute of Science, Bangalore, India\\Email: mechperson@gmail.com, \{sanandkr, vinod\}@ece.iisc.ernet.in}}
\maketitle 

\section*{Abstract}
\textbf{\textit{We consider nonparametric sequential hypothesis testing problem when the distribution
under the null hypothesis is fully known but the alternate hypothesis corresponds to some other unknown
distribution with some loose constraints. We propose a simple algorithm to address the problem. This is also generalized to the case when the distribution under the null hypothesis is not fully known. These problems are primarily motivated from wireless sensor networks and spectrum sensing in Cognitive Radios. A decentralized version utilizing spatial diversity is also proposed. Its performance is analysed and asymptotic properties are proved. The simulated and analysed performance of the algorithm are shown to be better than an earlier algorithm addressing the same problem with similar assumptions. We also modify the algorithm for optimizing performance when information about the prior probabilities of occurrence of the two hypotheses are known.}}
\\Key words: Sequential Detection, Distributed detection, Asymptotic performance.
\section{Introduction}
Presently there is a scarcity of spectrum due to the proliferation of wireless services. However, it has been observed that much of the licensed spectrum remains underutilised for most of the time. Cognitive Radios (CRs) are proposed as a solution to this problem (\cite{Akyildiz06nextgenerationdynamic}). These are designed to exploit the unutilised spectrum for their communication, without causing interference to the primary users. This is achieved through spectrum sensing by the CRs, to gain knowledge about spectrum usage by the primary users.
\par For CRs, spectrum sensing needs to be achieved at very low SNR in a wireless channel (\cite{Akyildiz06nextgenerationdynamic}). Distributed detection, which can mitigate the time-variance, fading, shadowing and electromagnetic interference, is very well-suited for this application. Thus distributed detection has been a highly-studied topic recently (\cite{Akyildiz06nextgenerationdynamic},\cite{vvv2007sensor}, \cite{varshney97sensors}). This has also found applications in sensor networks (\cite{vvv2007sensor}, \cite{varshney97sensors}). A relatively early treatment of distributed detection was given by \cite{tartakovsky}.
\par Distributed detection problems can be looked upon in a centralised or decentralised framework (\cite{vvv2007sensor}). In a centralised algorithm, the information collected by the local nodes are transmitted directly to the fusion centre which then uses it as a usual detection problem. In a decentralised algorithm, the local nodes transmit certain quantized values (or local decisions) to the fusion node. This has the advantage of requiring less power and bandwidth in transmission, but is suboptimal since the fusion centre has to take a decision based on less information. This kind of problems have been studied in \cite{vvv2007sensor}, \cite{fellouris}, \cite{sharmaenergyefficient} and \cite{mei}.
\par Distributed detection problems can also be classified as fixed sample size or sequential (\cite{Akyildiz06nextgenerationdynamic}, \cite{varshney97sensors}). In a fixed sample framework, the decision has to be made based on a fixed number of samples, and a likelihood ratio test turns out to be optimal for a simple binary hypothesis problem. In a sequential framework, samples are taken until some conditions are fulfilled, and once the process of taking samples has stopped, a decision is arrived at. It is known that in case of a single node, Sequential Probability Ratio Test (SPRT) outperforms other sequential or $\mbox{fiﬁxed}$ sample size detectors in the simple binary hypothesis problem (\cite{wald}). But optimal solutions in the decentralized setup are not available (\cite{basar}). In the parametric case, with full knowledge about the distributions, \cite{fellouris} and \cite{mei} propose an asymptotically optimal decentralized sequential test when the communication channel between the local nodes and FC is perfect.
\par For nonparametric sequential setup, \cite{mukhopadhyay} has provided separate algorithms for different problems like changes in mean, variance, etc. 
\par References \cite{dualsprtconf} and  \cite{uscslrtconf} have studied the distributed decentralised detection problem in a sequential framework, with a noisy reporting MAC. Algorithms in \cite{dualsprtconf} require the probability distributions involved from a paramteric family. The approach in \cite{uscslrtconf} is non-paramteric in the sense that it assumes very little knowledge of one of the distributions. In this paper, we present a simpler algorithm to address the problem studied in \cite{uscslrtconf}. Our algorithm has the added advantage of better performance in most cases, as borne out by simulations and analysis.
\par The paper is organized as follows. Section II presents the system model and the algorithm. Section III provides theoretical analysis of the algorithm for a single node case. Section IV provides an approximate analysis of the distributed algorithm. The asymptotics of the distributed algorithm are studied in Section V. Section VI compares our algorithm to KTSLRT in \cite{uscslrt}. Section VII provides a generalization of our algorithm along with an explanation in the CR setup. Section VIII concludes the paper.   
\section{System Model}
There are $L$ nodes and one fusion centre. Node $l$ makes observation $X_{k,l}$ at time $k$. We assume $\{X_{k,l}, k\geq 0\}$ are i.i.d. (independent identically distributed). We also assume that the observations received by different nodes are independent of each other. The distribution of the observations at each node is either $P_{0}$ or $P_{1}$. Each local node makes a decision based on the observations it receives and conveys the decision to the fusion node. The fusion node makes the final decision based on the local decisions it receives. The decision to be made is,\\
\\
$H_{0}$: if the probability distribution is $P_{0}$ and
\\
$H_{1}$: if the probability distribution is $P_{1}$.
\\
\\
We assume that $P_{0}$ is known, but that $P_{1}$ belongs to the family$\{P: D(P||P_{0}) \geq \lambda\mbox{ and }H(P)\geq H(P_{0})\}$, where $D(P_{1}||P_{0})$ is the divergence $E_{P_{1}}[\log\frac{P_{1}(X)}{P_{0}(X)}]$ and $H(P)$ is the entropy, or differential entropy, of the distribution $P$. The distribution $P_{1}$ can be different for different nodes, allowing for different fading gains. 
\par Our motivation for this setup is the Cognitive Radio (CR) system. A CR node has to detect if a channel is free (the primary node is not transmitting) or not. When the channel is free then the observations $X_{k,l}$ is the receiver noise with distribution $P_{0}$. This will often be known (even Gaussian) and hence it is reasonable to assume that $P_{0}$ is known (see, however, the generalization in Section VII). But when the primary is transmitting, it could be using adaptive modulation and coding, unknown to the secondary node, and even the fading of the wireless channel from the primary transmitter to the local CR node may be time-varying and not known to the receiver. This leads to an unknown distribution $P_{1}$ under $H_{1}$. We will elaborate in Section VII how this scenario can lead to the above class of distributions qualified for $H_{1}$. 
\par We have chosen sequential framework for detection at the local nodes as well as at the FC because it provides faster decisions on the average, for any probabilities of errors. 
\par Our detection algorithm works as follows. Each local node $l$, on receiving $X_{k,l}$ at time $k$, computes $W_{k,l}$ as 
\begin{equation}
W_{k,l}=W_{k-1,l}-\log P_{0}(X_{k,l}) - H(P_{0}) -\displaystyle\frac{\lambda}{2},\mbox{ }W_{0,l}=0.
\end{equation}  
If $W_{k,l}\geq-\log\alpha_{l}$, node $l$ decides $H_{1}$; if $W_{k,l}\leq\log\beta_{l}$, it decides $H_{0}$; otherwise it waits for the next observation. If at time $k$, node $l$ has decided $H_{0}$, it transmits this decision to the FC by transmiting $b_{0}$; if it has decided $H_{1}$ it transmits $b_{1}$; otherwise it sends nothing (i.e., 0). In the algorithm, $\alpha_{l}$, $\beta_{l}$, $b_{0}$ and $b_{1}$ are constants appropriately chosen so as to provide good system performance. 
\par Let $Y_{k,l}$ be the transmission from node $l$ to the FC at time $k$. The FC receives from local nodes at time $k$,
\begin{equation*}
Y_{k}=\displaystyle\sum_{l=1}^{L}Y_{k,l} + Z_{k},
\end{equation*}
where $Z_{k}$ is the FC receiver noise. We will assume $\{Z_{k}\}$ to be i.i.d. Because of $Z_{k}$, the FC does not directly know the local dcisions of the nodes. Thus it cannot use the majority rule, AND-rule, etc. usually used in literature. An advantage of allowing all nodes to transmit at the same time is that it reduces transmission delays from the nodes to the FC. 
\par The nodes keep transmitting to the FC till the FC makes its decision. Once the FC makes the decision, it will broadcast a message to all the local nodes to stop transmission. 
\par The local nodes make their decisions at random times. Thus $\{Y_{k}\}$ received by the FC are not i.i.d. However, inspired by sequential detection algorithms (e.g. SPRT as in \cite{wald}) it uses the following algorithm to make decisions. At time $k$, it computes \begin{equation*}F_{k}=F_{k-1}+\displaystyle\log\frac{g_{\mu_{1}}(Y_{k})}{g_{\mu_{0}}(Y_{k})},\mbox{ }F_{0}=0.\end{equation*}
\par After that, it decides $H_{0}$ if $F_{k}\leq\log\beta$; $H_{1}$ if $F_{k}\geq\log\alpha$ and waits for the next observation otherwise. Here the distributions $g_{\mu_{0}}$ and $g_{\mu_{1}}$ are appropriately decided. We assume that the distribution of noise $\{Z_{k}\}$ is known. Then $g_{\mu_{0}}$ is the distribution of $\mu_{0}+Z_{k}$, and $g_{\mu_{1}}$ is the distribution of $\mu_{1}+Z_{k}$, where $\mu_{0}$ and $\mu_{1}$ are constants. By choosing $\mu_{0}$ and $\mu_{1}$ appropriately, we can ensure that FC makes a decision which is (say) close to a majority decision of local nodes.
\par The overall algorithm is summarized as
\begin{enumerate}[label={(\roman*)}]
\item Node $l$ receives $X_{k,l}$ at time $k \geq 1$ and computes $W_{k,l}$.
\item Node $l$ transmits $Y_{k,l} = b_{1}\mathbbm{1}\{W_{k,l} \geq -\log\alpha_{l}\} + b_{0}\mathbbm{1}\{W_{k,l}\leq \log\beta_{l}\}$.
\item Fusion node receives at time $k$,
\begin{equation*}Y_{k} = \displaystyle \sum_{l=1}^{L} Y_{k,l} + Z_{k}.\end{equation*}
\item Fusion node computes, 
\begin{equation*}F_{k} = F_{k-1} + \log\frac{g_{\mu_{1}}(Y_{k})}{g_{\mu_{0}}(Y_{k})},\mbox{ }F_{0} = 0.\end{equation*}
\item Fusion node decides $H_{0}$ if $F_{k} \leq \log\beta$, $H_{1}$ if $F_{k} \geq -\log\alpha$; otherwise it waits for the next observation. 
\end{enumerate}
 \par In the rest of the paper we analyze the performance of this algorithm. First we analyze the performance for a single (local) node in Section III and then for the decentralised algorithm in Section IV. Proofs of the lemmas and theorems provided are similar to those in \cite{dualsprt} and \cite{uscslrt} and are skipped in this paper for lack of space. We will show that the present algorithm provides a better performance than \cite{uscslrt}. Also, unlike in \cite{uscslrt}, it is simpler to implement because it does not require a universal source coder and it also does not require quantization of observations. 
\par In the following, $E_{i}[.]$ and $P_{i}(.)$ denote the expectation and probability, respectively, under $H_{i}$, $i=0,1$. 
\section{Performance Analysis for a Single Node}
In this section we provide performance of the test for a single node. We will omit the node index $l$ in this section. Let
\begin{equation*}N_{1}\triangleq \inf\{n:\mbox{ }W_{n} >-\log \alpha\},\end{equation*}
\begin{equation*}N_{0}\triangleq \inf\{n:\mbox{ }W_{n} <\log \beta\},\end{equation*}
\begin{equation*}N\triangleq \min(N_{0},N_{1}).\end{equation*}
$\mathbf{Lemma\mbox{ }3.1}$.   $P(N<\infty)=1$ under $H_{0}$ and $H_{1}$.
                  \hfill $\blacksquare$\\ 
\par We will use the notation that $P_{FA}\triangleq P_{0}(\mbox{ decide }H_{1})$ and $P_{MD}\triangleq P_{1}(\mbox{ decide }H_{0})$. Also, $f(x)=\mathcal{O}[g(x)]$ denotes that $\displaystyle\limsup_{x\to\infty}\frac{f(x)}{g(x)}<\infty$.\\ \\ 
$\mathbf{Theorem\mbox{ }3.2}$.   a) $P_{FA}=\mathcal{O}(\alpha^{s})$ where $s$ is a solution of \begin{equation*}\displaystyle E_{0}[e^{-s(H(P_{0})+\log P_{0}(X_{k})+\frac{\lambda}{2}-\epsilon)}]=1,\end{equation*} $0<\epsilon<\frac{\lambda}{2}$ and $s>0$.\\ b) $P_{MD}=\mathcal{O}(\beta^{s^{*}})$ where $s^{*}$ is a solution of\begin{equation*}E_{1}[e^{-s^{*}(-H(P_{0})-\log P_{0}(X_{k})-\frac{\lambda}{2}+\epsilon)}]=1,\end{equation*} $0<\epsilon<D(P_{1}||P_{0})+H(P_{1})-H(P_{0})-\frac{\lambda}{2}$ and $s^{*}>0.$ \hfill $\blacksquare$ \\ 
\par From \cite{quickestdetection}, pg 78-79, $s$ and $s^{*}$ defined in Theorem 3.2 exist.\\
$\mathbf{Theorem\mbox{ }3.3}$:  a) Under $H_{0}$, $\displaystyle \lim_{\alpha,\beta\to 0}\frac{N}{|\log\beta|}=\frac{2}{\lambda}$ a.s.\\ 
If in addition, $E_{0}[|(\log P_{0}(X))|^{p}]<\infty$ for some $p\geq 1$, then\begin{equation*} \displaystyle \lim_{\alpha,\beta\to 0}\frac{E_{0}[N^{q}]}{|\log\beta|^{q}}=(\frac{2}{\lambda})^{q},\end{equation*}for all $0<q\leq p$.\\ 
b)  Under $H_{1}$,\begin{equation*}\displaystyle \lim_{\alpha,\beta\to 0}\frac{N}{|\log\beta|}=\frac{1}{D(P_{1}||P_{0})+H(P_{1})-H(P_{0})-\frac{\lambda}{2}}\mbox{ a.s.}\end{equation*} 
If in addition, $E_{1}[|(\log P_{0}(X))|^{p}]<\infty$ for some $p\geq 1$, then\begin{equation*}\displaystyle \lim_{\alpha,\beta\to 0}\frac{E_{1}[N^{q}]}{|\log\beta|^{q}} = (D(P_{1}||P_{0})+H(P_{1})-H(P_{0})-\frac{\lambda}{2})^{-q},\end{equation*} for all $0<q\leq p$. \hfill $\blacksquare$
\\
\par From Lemma 3.1 and Theorem 3.2 we see that the asymptotic behaviour of our algorithm is comparable to that in \cite{uscslrt} and also to dualSPRT (\cite{dualsprt}). However, we will see via simulations that it substantially outperforms KTSLRT in \cite{uscslrt}. 
\section{Approximate Performance of Decentralised Algorithm}
In the following, we take, for convenience, $\alpha_{l}=\beta_{l}\mbox{  } \forall l$, $\alpha=\beta$, $b_{1}=-b_{0} = b$, and $\mu_{1} = -\mu_{0} = \mu = I.b$, for some $1\leq I\leq L$. Roughly speaking, this ensures that the FC makes decisions $H_{1}$ when $I$ more nodes decide $H_{1}$ compared to the nodes deciding $H_{0}$. Similarly for $H_{0}$. In the following, $N_{l}^{i}$ corresponds to $N^{i}$ at node $l$ and $N_{l}$ corresponds to $N$. Similarly, $N^{1}$, $N^{0}$ and $N$ represent the corresponding terms for the FC. 
\\
\\
$\mathbf{Lemma\mbox{ }4.1}$.  For $i=0$, $1$,
\begin{equation*}P_{i}(N_{l}=N_{l}^{i}) \to 1\mbox{ as }\alpha_{l}, \beta_{l}\to 0,\end{equation*} 
\begin{equation*}P_{i}(N=N^{i}) \to 1\mbox{ as }\alpha_{l}, \beta_{l}\to 0\mbox{ and }\alpha, \beta\to 0.\hfill \mbox{                   
             }\blacksquare\end{equation*}  
Note: In general, when $\alpha_{l}\neq\beta_{l}$, the results of Lemma 4.1 under $H_{0}$ demand that $\beta$ and/or $\beta_{l}\to 0$, and the results under $H_{1}$ demand that $\alpha$ and/or $\alpha_{l}\to 0$. Analogous comments will hold for the subsequent results as well.  
\\
\\$\mathbf{Lemma\mbox{ }4.2}$.  Under $H_{i}$,\\
a) $|N_{l}-N_{l}^{i}| \to 0$ a.s. as $\alpha_{l}$, $\beta_{l}\to 0$ and \begin{equation*}\displaystyle \lim_{\alpha_{l} \to 0}\frac{N_{l}}{|\log \alpha_{l}|} = \lim_{\alpha_{l} \to 0}\frac{N_{l}^{i}}{|\log \alpha_{l}|} = \frac{1}{|\delta_{i,l}|}\mbox{ a.s. and in }L^{1}.\end{equation*} 
b)  $|N-N^{i}| \to 0$ a.s. and \begin{equation*}\displaystyle \lim \frac{N}{|\log \alpha|} = \lim \frac{N^{i}}{|\log \alpha|}\mbox{ a.s. and in }L^{1},\end{equation*}\\ as $\alpha_{l}, \beta_{l} \to 0\mbox{ and }\alpha, \beta \to 0$.\hfill $\blacksquare$
\\
\par Lemmas 4.1 and 4.2 show that the local nodes make the right decisions as the thresholds $|\log\alpha_{l}|$ and $|\log\beta_{l}|$ tend to infinity. Then the FC also makes the right decisions when its own thresholds increase. We need to set the thresholds such that the probabilities of errors are small.\\ 
\par We will use the following notation:\\
$\delta_{i,FC}^{j}\triangleq$ mean drift of the fusion centre process $\{F_{k}\}$ under $H_{i}$, when $j$ local nodes are transmitting.\\
$t_{j}\triangleq$ time at which the mean drift of $\{F_{k}\}$ changes from $\delta_{i,FC}^{j-1}$ to $\delta_{i,FC}^{j}$.\\ \\
$\tilde{F}_{j}\triangleq E[F_{t_{j}-1}]$
\\
Now, it is seen that under $H_{i}$,\begin{equation*}\tilde{F}_{j} = \tilde{F}_{j-1} + \delta^{j-1}_{i, FC}(E(t_{j}) - E(t_{j-1})), \tilde{F}_{0} = 0.\end{equation*}
$\mathbf{Lemma\mbox{ }4.3}$.  $P_{i}(\mbox{decision of local node }l\mbox{ at time }t_{k}\mbox{ is }H_{i}$\\$\mbox{ and }t_{k}\mbox{ is the }k^{th}\mbox{ order statistics of }\{N_{1}^{i}\mbox{, ... }N_{L}^{i}\}) \to 1$ \\as $\alpha_{l} \to 0,\mbox{ }\forall \mbox{ }l$. \hfill$\blacksquare$\\ \\
$\mathbf{Lemma\mbox{ }4.4}$.  Under $H_{0}$, when $\alpha_{l}$ and $\beta_{l}$ are small, 
\begin{equation*}\displaystyle N_{l}^{0} \sim \mathcal{N}(\frac{-|\log \alpha_{l}|}{\delta_{0,l}},\frac{-|\log \alpha_{l}|\rho_{0,l}^{2}}{\delta_{0,l}^{3}}),\end{equation*}
\\
where \begin{equation*}\delta_{0,l}\displaystyle\triangleq E_{0}[-\log P_{0}(X_{k,l}-H(P_{0})-\frac{\lambda}{2})],\mbox{ and}\end{equation*}\begin{equation*}\rho_{0,l}^{2}\displaystyle\triangleq\mbox{ variance of }[-\log P_{0}(X_{k,l})-H(P_{0})-\frac{\lambda}{2})]\mbox{ under }H_{0}.\end{equation*} 
\textbf{Proof}: See Theorem 5.1, Chapter 3 in \cite{opac-b1086480}.\hfill $\blacksquare$ \\A similar result holds for $H_{1}$ as well.\\
\par Let $E_{DD} \triangleq E_{i}[N]$, $i=0,1$. In the following we provide an approximation for this for both $i=0,1$. 

\par When $\alpha_{l}$ and $\alpha$ are small, probabilities of error are small, as proved in the above lemmas. Hence in such a scenario, for approximation, we assume that local nodes are making correct decisions. 
\par Let,\\
$l_{0}^{*}\triangleq\min \{j: \delta^{j}_{0, FC} > 0 \mbox{ and }\frac{-|\log\alpha| - \tilde{F_{j}}}{\delta^{j}_{0, FC}}< E(t_{j+1}) - E(t_{j})\}$. \\ \\The detection delay $E_{DD}$ under $H_{0}$ can be approximated as,
\begin{equation}
E_{DD} \approx E(t_{l_{0}^{*}}) + \frac{-|\log \alpha| - \tilde{F}_{l_{0}^{*}}}{\delta^{l_{0}^{*}}_{0, FC}}.
\end{equation}
\par The first term in  approximation (2) corresponds to the mean time till the mean drift of $\{F_{k}\}$ becomes positive (for $H_{1}$) or negative (for $H_{0}$), and the second term corresponds to the mean time from then on till it crosses the threshold. Using the Gaussian approximation of Lemma 4.4, the $t_{k}$'s are the order statistics of i.i.d. Gaussian random variables and hence, the $\tilde{F}_{k}$'s can be computed. See, for example, \cite{gaussiannonidentical}. A similar approximation can be written for $E_{DD}$ under $H_{1}$ as well.
\par In the following, we compute approximate expressions for $P_{FA}\triangleq P[\mbox{decision is }H_{1}|H_{0}]$ and $P_{MD}\triangleq P[\mbox{decision is }H_{0}|H_{1}]$.
\par Under the same setup of small $\alpha_{l}$ and $\alpha$, for $P_{FA}$ analysis, we assume that all local nodes are making correct decisions. Then for false alarm, the dominant event is $\{N^{1}<t_{1}\}$. Also, for reasonable performance, $P_{0}(N^{0}<t_{1})$ should be small. Then, the probability of false alarm, $P_{FA}$, can be approximated as
\begin{equation}
P_{FA} = P_{0}(N^{1}<N^{0}) \geq P_{0}(N^{1}<t_{1},N^{0}>t_{1})\\ \approx P_{0}(N^{1}<t_{1}).
\end{equation}
Also,\\ \\$P_{0}(N^{1}<N^{0})\leq P_{0}(N^{1}<\infty)$\\$=P_{0}(N^{1}<t_{1})+P_{0}(t_{1}\leq N^{1}<t_{2})+\cdots$ \hfill (4)\\ 
The first term in the RHS of (4) should be the dominant term since after $t_{1}$, the drift of $F_{k}$ will have the desired sign (will at least be in the favourable direction) with a high probability, if the local nodes make correct decisions.
\par Equations (3) and (4) suggest that $P_{0}(N^{1}<t_{1})$ should serve as a good approximation for $P_{FA}$. Similar arguments show that $P_{1}(N^{0}<t_{1})$ should serve as a good approximation for $P_{MD}$. In the following, we provide approximations for these. 
\par Let $\xi_{k}\triangleq \displaystyle\log\frac{g_{\mu_{1}}(Y_{k})}{g_{\mu_{0}}(Y_{k})}$ before $t_{1}$ have mean 0 and probability distribution symmetric about 0. Then, from the Markov property of the random walk $\{F_{k}\}$, before $t_{1}$,
\\
\\
$P_{0}(N^{1}<t_{1})\approx$\\ $\displaystyle \sum_{k=1}^{\infty} P_{0}[\{F_{k}\geq -\log \alpha\}\bigcap_{n=1}^{k-1}\{F_{n} < -\log \alpha\}|t_{1}>k].P_{0}(t_{1}>k)$
\\
\\
$=\displaystyle \sum_{k=1}^{\infty} P_{0}[\{F_{k}\geq -\log \alpha\}|\bigcap_{n=1}^{k-1}\{F_{n} < -\log \alpha\}].P_{0}[\bigcap_{n=1}^{k-1}\{F_{n} < -\log \alpha\}].P_{0}(t_{1}>k)$\\ \\
$=\displaystyle \sum_{k=1}^{\infty} P_{0}(F_{k}\geq -\log \alpha|(F_{k-1} < -\log \alpha).P_{0}(\sup_{1 \leq n \leq k-1}F_{n} < -\log \alpha).[1-\Phi_{t_{1}}(k)]$
\\
$=\displaystyle \sum_{k=1}^{\infty} [\int_{u=0}^{\infty} P_{0}(\xi_{k}>u)\mbox{f}_{F_{k-1}}(-\log \alpha - u)du].P_{0}(\sup_{1\leq n\leq k-1}F_{n} < -\log \alpha).[1-\Phi_{t_{1}}(k)]$, where $\Phi_{t_{1}}$ is the CDF of $t_{1}$.
\\
\\
We can find a lower bound to the above expression by using\\ $P_{0}(\displaystyle \sup_{1\leq n\leq k-1}F_{n} < -\log \alpha) \geq 1-2P_{0}(F_{k-1}\geq-\log \alpha)$\\ \\(\cite{billingsley}, pg 525) and an upper bound by replacing $\displaystyle \sup_{1\leq n\leq k-1}F_{n}$ by $F_{k-1}$. 
\par Similarly, $P_{MD}$ can be approximated as\\
\\
$P_{MD}\gtrsim$\\ \\$\displaystyle \sum_{k=1}^{\infty} [\int_{u=0}^{\infty} P_{1}(\xi_{k}<-u)\mbox{f}_{F_{k-1}}(\log \beta + u)du].[1-2P_{1}(F_{k-1} \leq \log \beta)].[1-\Phi_{t_{1}}(k)]$, and 
\\ 
\\
$P_{MD}\lesssim$\\ \\$\displaystyle \sum_{k=1}^{\infty} [\int_{u=0}^{\infty} P_{1}(\xi_{k}<-u)\mbox{f}_{F_{k-1}}(\log \beta + u)du].P_{1}(F_{k-1} > \log \beta).[1-\Phi_{t_{1}}(k)]$.\\ In the above expressions, $\mbox{f}_{F_{k-1}}$ stands for the probability density function of $F_{k-1}$.
\par We will show in Section VI that these approximate results compare well with simulations.\\
\section{Asymptotic Results}
In this section, the following assumptions will be required for some of the results.
\begin{enumerate}[label={(\roman*)}]
	\item $E_{i}[V_{k,l}]^{p+1}<\infty$, for some $p > 1$, where $V_{k,l} =-\displaystyle\log P_{0}(X_{k,l})$.
	\item $E_{i}[|\xi_{k}^{*}|^{p+1}] < \infty$, where $\xi_{k}^{*}$ is the drift of $\{F_{k}\}$ when all nodes are transmitting wrong decisions.
\end{enumerate}
We will use the notation
\begin{enumerate}[label={}]
          \item $\theta_{i} = E_{i}[\xi_{k}^{*}]$,
          \item $\mathcal{A}^{i} = \{\omega \in \Omega : \mbox{all local nodes transmit correct decisions }$\\$(b_{i})\mbox{ under } H_{i} \}$,
          \item $\Delta(\mathcal{A}^{i}) = $ mean drift of fusion centre LLR under $\mathcal{A}^{i}$, i.e., $E_{i}[\xi_{k}|\mathcal{A}^{i}]$,
          \item $D^{0}_{tot} = \frac{L\lambda}{2}$,
          \item $D^{1}_{tot}  = \displaystyle \sum_{l=1}^{L}[D(f_{1,l}||f_{0,l})+H(f_{1,l}) - H(f_{0,l}) - \frac{\lambda}{2}]$,
	\item $r_{l} = \frac{1}{L}$,
 	\item $\rho_{l}= \frac{D(f_{1,l}||f_{0,l})+H(f_{1,l}) - H(f_{0,l}) - \frac{\lambda}{2}}{D^{1}_{tot}}$,
           \item $\Lambda_{i}(\alpha)=\displaystyle \sup_{\lambda}[a\lambda - \log g_{i}(\lambda)]$,
           \item $g_{i} =$ moment-generating function of $|\xi_{k}^{*}|$,
           \item $\alpha_{i}^{+} =\mbox{ess}\sup |\xi_{k}^{*}|$.
\end{enumerate}
Furthermore, local node thresholds are $-r_{l}|\log c|$ and $\rho_{l}|\log c|$, where c is a constant, and the fusion centre thresholds are $-|\log c|$ and $|\log c|$.\\ \\
$\mathbf{Theorem\mbox{ 5.1}}$.   Under$H_{i}$,\begin{equation}\displaystyle \limsup_{c\to 0}\frac{N}{|\log c|} \leq \frac{1}{D^{i}_{tot}} + \frac{C_{i}}{\Delta(\mathcal{A}^{i})}\end{equation} a.s. and (if conditions (i) and (ii) above hold) also in $L^{p}$, where $C_{0}=-(1+\frac{\theta_{0}}{D^{0}_{tot}})$ and $C_{1}=1+\frac{\theta_{1}}{D^{1}_{tot}}$.   \hfill$\blacksquare$\\
\par We introduce the following function, \\ $s_{i}(\eta) \triangleq 
\begin{cases}
\frac{\eta}{\alpha_{i}^{+}}, & \text{if }\eta \geq \Lambda_{i}(\alpha_{i}^{+})\\
\frac{\eta}{\Lambda_{i}^{-1}(\eta)}, & \text{if } \eta \in (0,\Lambda_{i}(\alpha_{i}^{+}))
\end{cases}$,\\ \\
where $g_{i}(\lambda)=$ m.g.f. of $|\xi_{k}^{*}|$, and $\Lambda_{i}(\alpha) = \displaystyle \sup_{\lambda}[a\lambda - \log g_{i}(\lambda)]$.\\
$R_{i}\triangleq$\\$\displaystyle \min_{1 \leq l \leq L} \{-\log \inf_{t \geq 0} E_{i}[\exp\{-t \log\frac{f_{1,l}(X_{k,l})}{f_{0,l}(X_{k,l})}\}]\}$.
\\ \\
$\mathbf{Theorem\mbox{ 5.2}}$.   Let $g_{i}(\lambda)<\infty$ in a neighbourhood of zero. Then, the following hold: 
\\
a)  $\displaystyle \lim_{c\to 0}\frac{P_{FA}}{c} = 0$ if for some $0<\eta<R_{0}$, $s_{0}(\eta) > 1$.\\ \\
b)  $\displaystyle \lim_{c\to 0}\frac{P_{MD}}{c} = 0$ if for some $0<\eta<R_{1}$, $s_{1}(\eta) > 1$.
$\blacksquare$
\par As a specific example, it can be easily shown that under the fusion conditions assumed for the simulations in the next section, $s_{i}(\eta)>1\mbox{ for }\eta\approx 0.0335$, for $i=0,\mbox{ }1$. Hence for $R_{0}, R_{1}\geq 0.0335$, the conditions of Theorem 5.2 hold. If the distributions are $\mathcal{N}(0,\sigma^{2})$ and $\mathcal{N}(\mu,\sigma^{2})$ under $H_{0}$ and $H_{1}$ respectively, or $\ln\mathcal{N}(0,\sigma^{2})$ and $\ln\mathcal{N}(\mu,\sigma^{2})$ under $H_{0}$ and $H_{1}$ respectively (two of the situations taken for simulations), it is seen that $R_{0}=0$ and $R_{1}=\displaystyle\frac{\mu^{2}}{8\sigma^{2}}$. Hence, even though the condition for $P_{FA}$ is not satisfied, the condition for $P_{MD}$ will be satisfied if $\mu\geq 0.4977\sigma$. Taking $\sigma^{2}=1$, this gives $\mu\geq -6.06 dB$, which is satisfied in the Gaussian example considered in the next section. Also, taking $\sigma^{2}=3$, this gives $\mu\geq 1.4931$, which is satisfied in the Lognormal example considered in the next section. 
\par From Theorems 5.1 and 5.2 we see that the asymptotic performance of our algorithm is comparable to SPRT and KTSLRT in \cite{dualsprt} and \cite{uscslrt}.
\section{Simulations}
In this section, we compare the simulated and theoretical performances of the new algorithm with KT-SLRT (\cite{uscslrt}). For simulations, we have taken $b_{1}=-b_{0}=1$, $L=5$, $\mu_{1}=-\mu_{0}=2$. Also, the FC noise has been taken as zero mean Gaussian with variance $\sigma^{2}$. Hence in this case, $\Delta(\mathcal{A}^{1})$ $=$ $-\Delta(\mathcal{A}^{0})$ $=$ $\frac{20}{\sigma^{2}}$ $=$ $\theta_{0}$ $=$ $\theta_{1}$. In the following simulations, $E_{DD}\triangleq 0.5[E_{0}(N)+E_{1}(N)]$ and $P_{e}\triangleq 0.5(P_{FA}+P_{MD})$
\par The observations $X_{k,l}$ are considered with the following distributions: \\
Pareto Distribution $\mathcal{P}$, $P_{0} \sim  \mathcal{P}(10,2)$ and $P_{1} \sim  \mathcal{P}(3,2)$.
\\
Lognormal Distribution $\ln\mathcal{N}$, $P_{0} \sim  \ln \mathcal{N}(0,3)$ and $P_{1} \sim  \ln \mathcal{N}(3,3)$.
\\
Gaussian Distribution $\mathcal{N}$ with $P_{0} \sim  \mathcal{N}(0,1)$ and $P_{1} \sim\mathcal{N}(1,1)$. \\The channel gains from the primary to the secondary nodes are 1 except in the Gaussian case, where these are taken as 0 dB, -1.5 dB, -2.5 dB, -4 dB and -6 dB.
\par We plot the results in Figs 1-7. We see that the new algorithm markedly outperforms KT-SLRT. This may be due to the presence of compression in KT-SLRT, due to which redundancy is introduced, leading to inaccuracies in the estimate. Also, the approximations provided in Section IV are much closer to the simulated values than the asymptotics.\\ \\
\begin{figure}[h]
        \vspace{-20pt}
        \centering
        \begin{subfigure}[b]{0.45\textwidth}
                \centering
                \includegraphics[width=\textwidth,height=4cm]{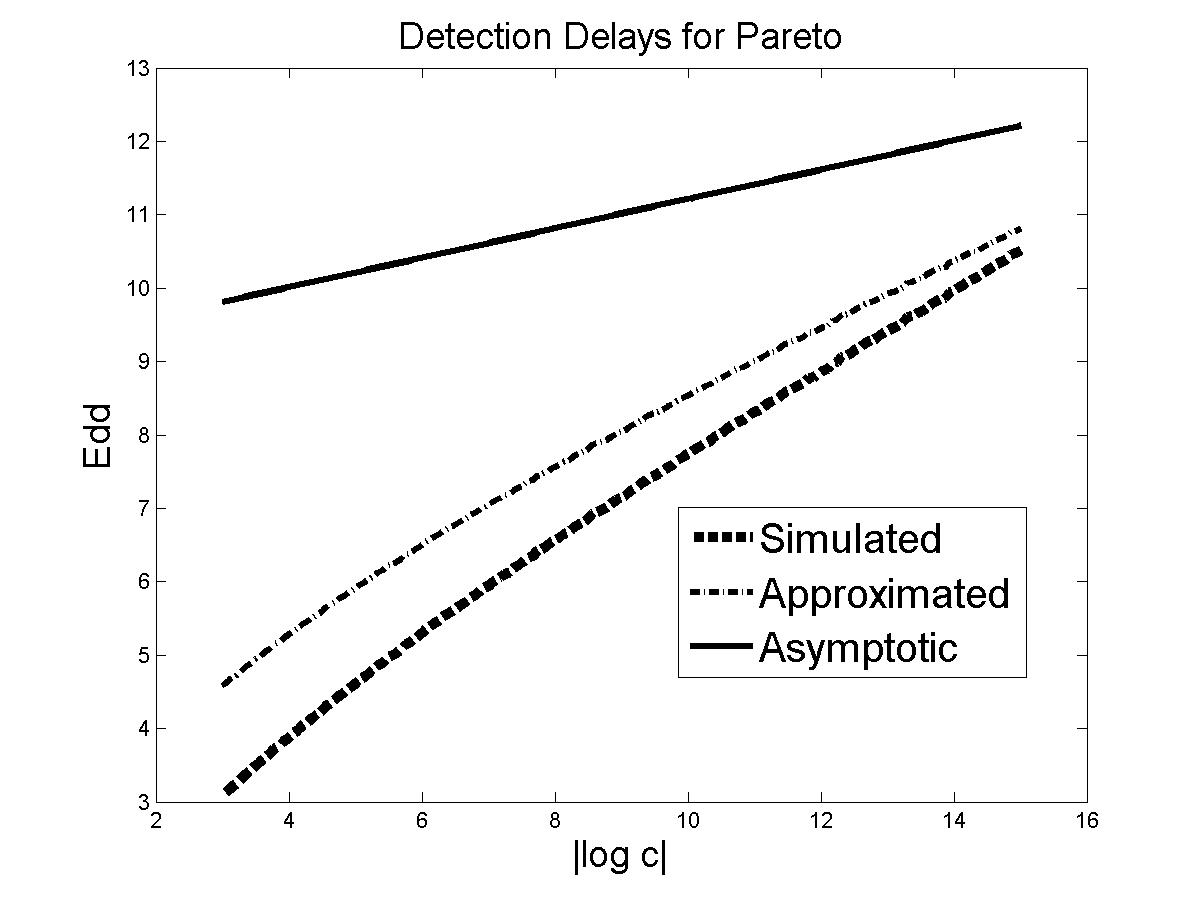}
                \label{fig:gull}
        \end{subfigure}
        \vspace{-20pt}
        \begin{subfigure}[b]{0.45\textwidth}
                \centering
                \includegraphics[width=\textwidth,height=4cm]{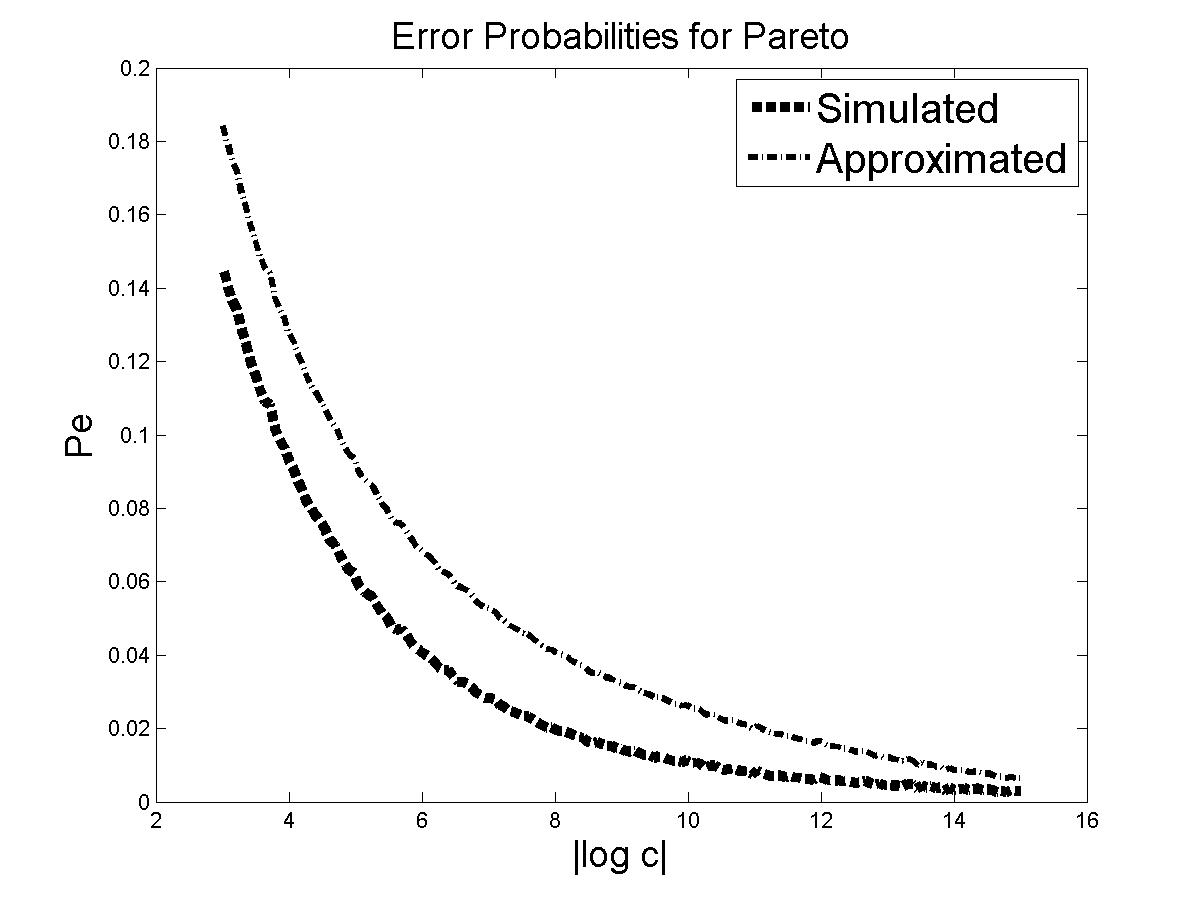}
                \label{fig:tiger}
        \end{subfigure}
        \caption{Performance for Pareto Distribution\\ $\mbox{       }$Top: Detection Delay; Bottom: Error Rate}\label{fig:animals}
\end{figure}
\begin{figure}[h]
        \vspace{-20pt}
        \centering
        \begin{subfigure}[b]{0.45\textwidth}
                \centering
                \includegraphics[width=\textwidth,height=4cm]{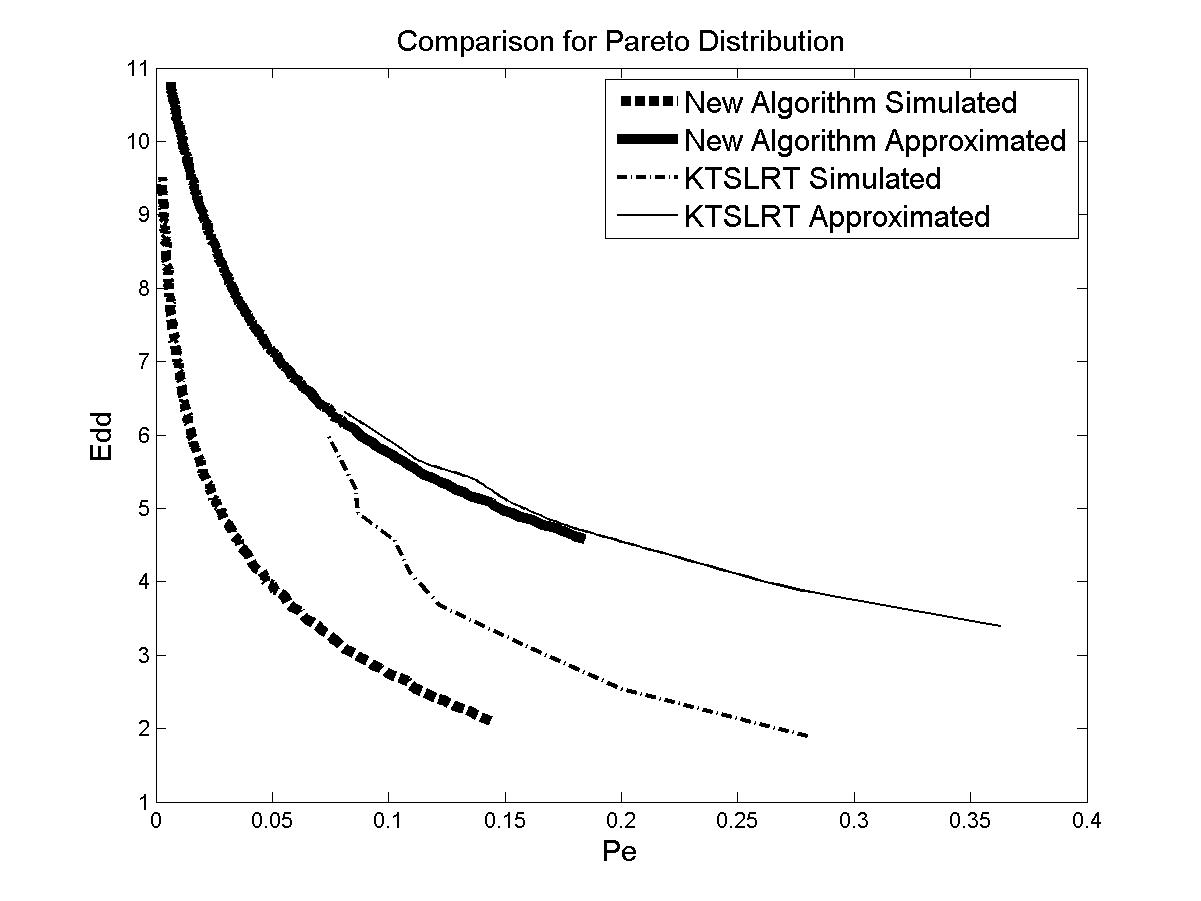}
                \label{fig:gull}
        \end{subfigure}
        \caption{Performance Comparison with KT-SLRT for Pareto Distribution}\label{fig:animals}
\end{figure}
\begin{figure}[h]
        \centering
        \begin{subfigure}[b]{0.45\textwidth}
                \centering
                \includegraphics[width=\textwidth,height=4cm]{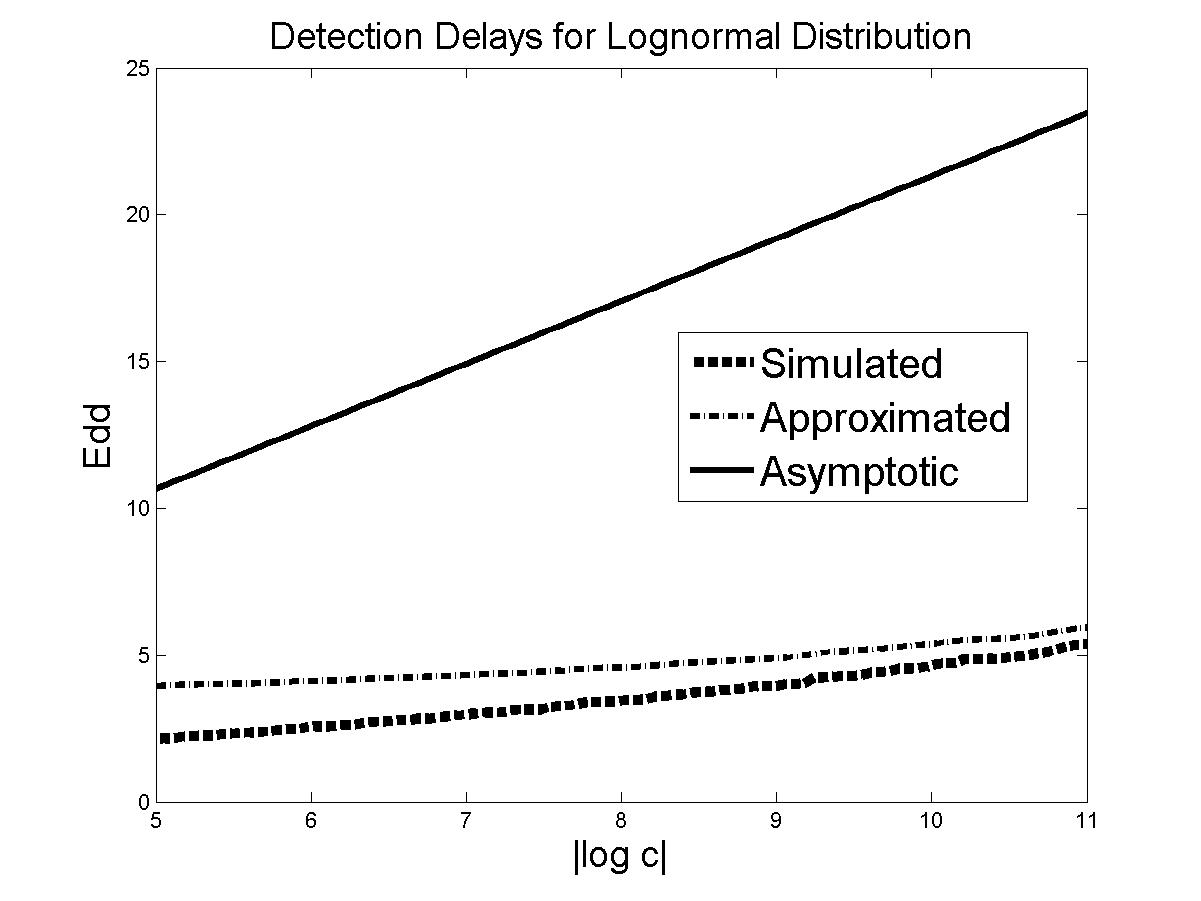}
                \label{fig:gull}
        \end{subfigure}
        \vspace{-20pt}
        \begin{subfigure}[b]{0.45\textwidth}
                \centering
                \includegraphics[width=\textwidth,height=4cm]{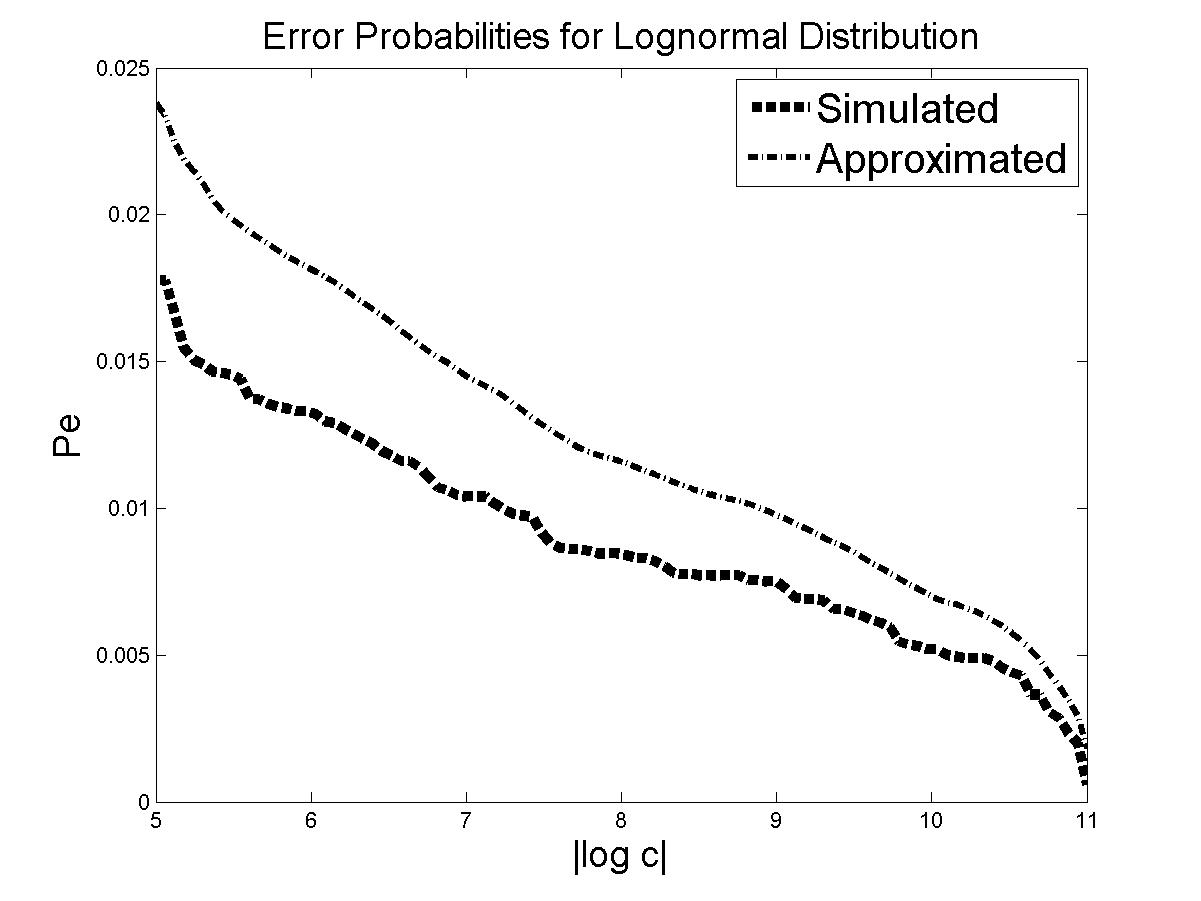}
                \label{fig:tiger}
        \end{subfigure}
        \caption{Performance for Lognormal Distribution\\ $\mbox{       }$Top: Detection Delay; Bottom: Error Rate}\label{fig:animals}
\end{figure}
\begin{figure}[h]
        \centering
        \begin{subfigure}[b]{0.45\textwidth}
                \centering
                \includegraphics[width=\textwidth,height=4cm]{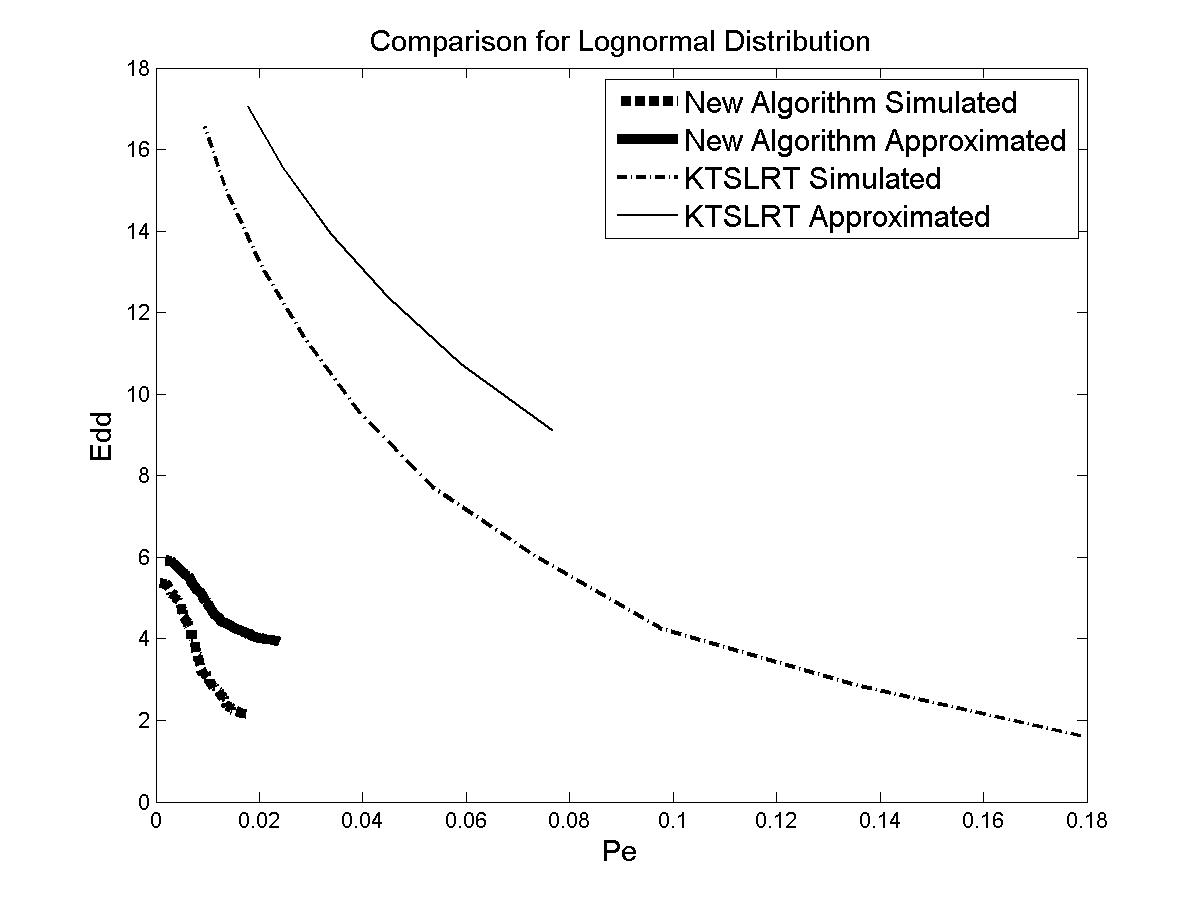}
                \label{fig:gull}
        \end{subfigure}
        \caption{Performance Comparison with KT-SLRT for Lognormal Distribution}\label{fig:animals}
\end{figure}
\begin{figure}[h]
        \centering
        \begin{subfigure}[b]{0.45\textwidth}
                \centering
                \includegraphics[width=\textwidth,height=4cm]{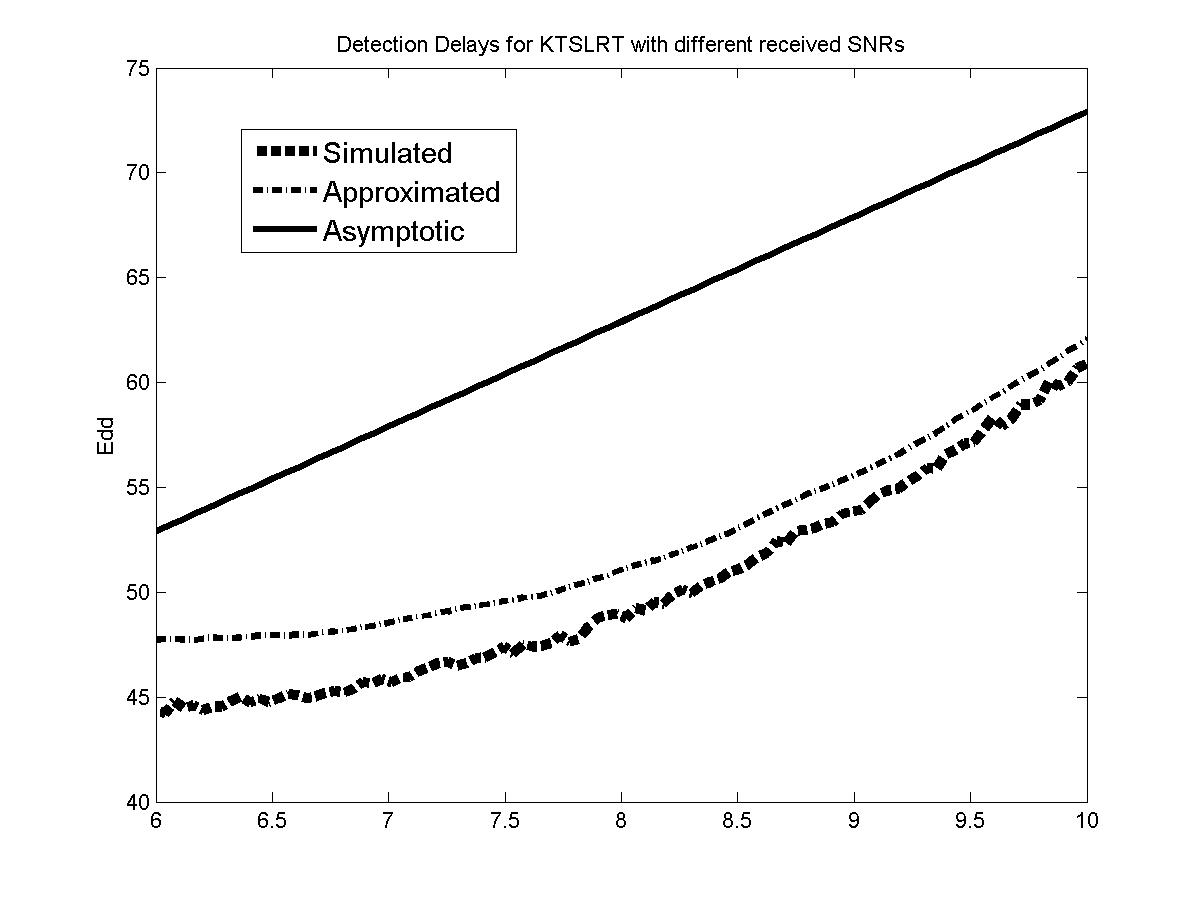}
                \label{fig:gull}
        \end{subfigure}
        \vspace{-20pt}
        \begin{subfigure}[b]{0.45\textwidth}
                \centering
                \includegraphics[width=\textwidth,height=4cm]{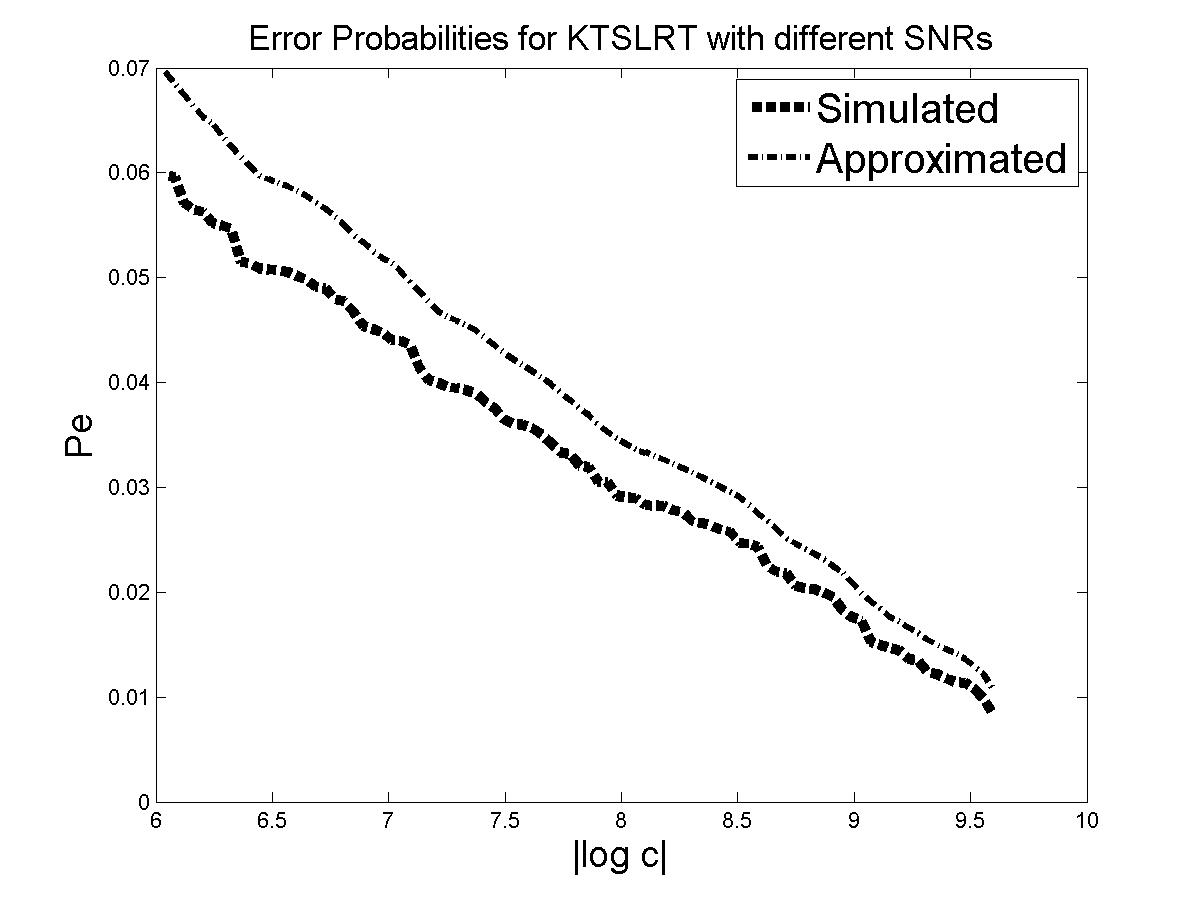}
                \label{fig:tiger}
        \end{subfigure}
        \caption{Performance of KTSLRT for Gaussian Distribution with Different Received SNRs\\ $\mbox{       }$Top: Detection Delay; Bottom: Error Rate}\label{fig:animals}
\end{figure}
\begin{figure}[h]
        \centering
        \begin{subfigure}[b]{0.45\textwidth}
                \centering
                \includegraphics[width=\textwidth,height=4cm]{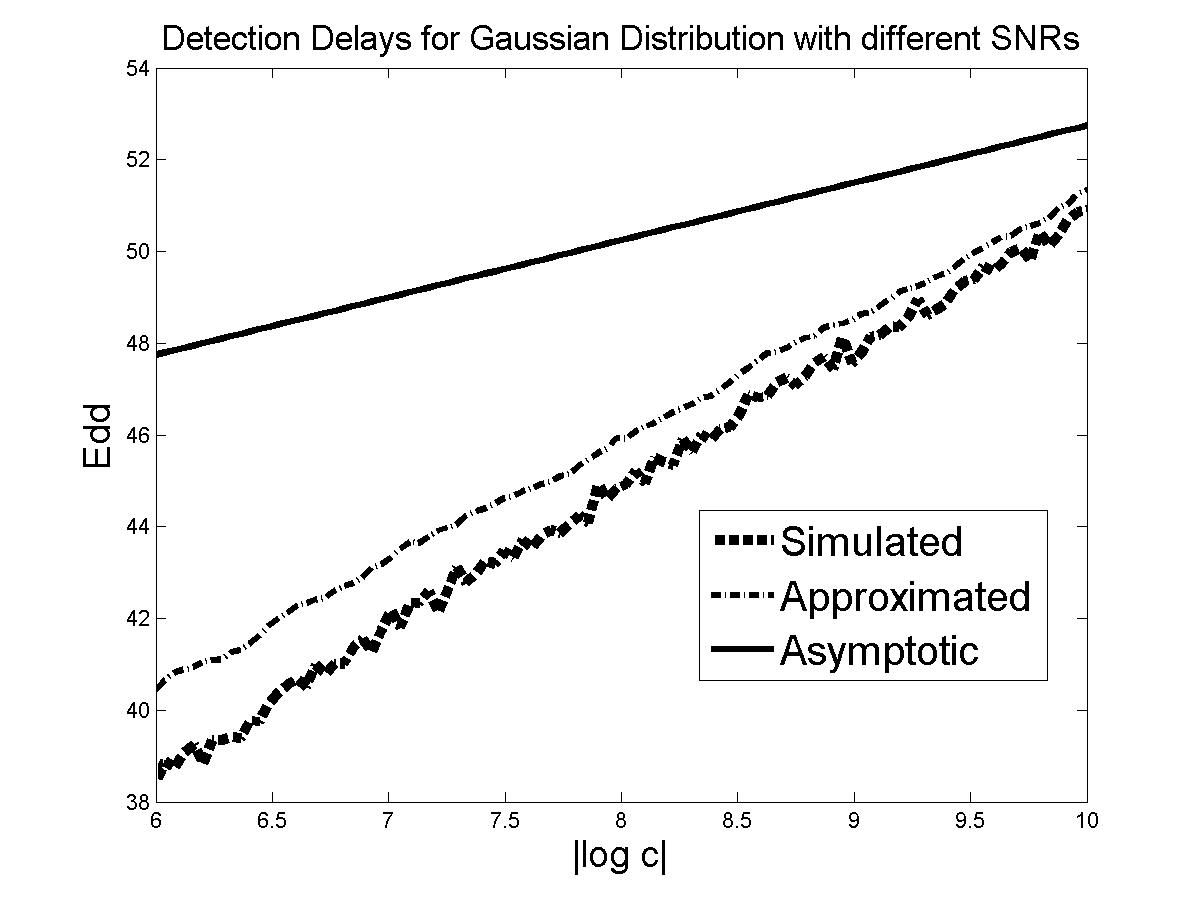}
                \label{fig:gull}
        \end{subfigure}
        \vspace{-20pt}
        \begin{subfigure}[b]{0.45\textwidth}
                \centering
                \includegraphics[width=\textwidth,height=4cm]{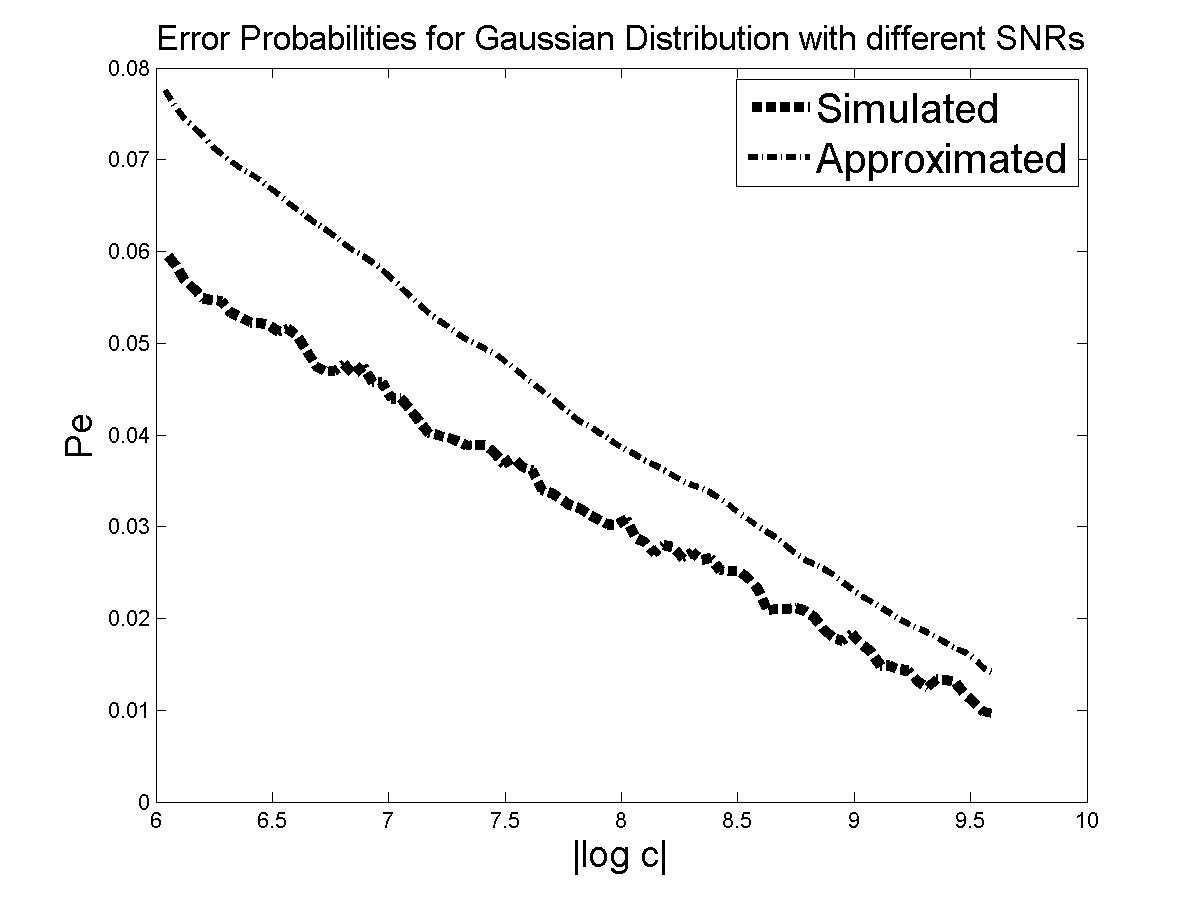}
                \label{fig:tiger}
        \end{subfigure}
        \caption{Performance for Gaussian Distribution with Different Received SNRs\\ $\mbox{       }$Top: Detection Delay; Bottom: Error Rate}\label{fig:animals}
\end{figure}
\begin{figure}[h]
        \centering
        \begin{subfigure}[b]{0.45\textwidth}
                \centering
                \includegraphics[width=\textwidth,height=4cm]{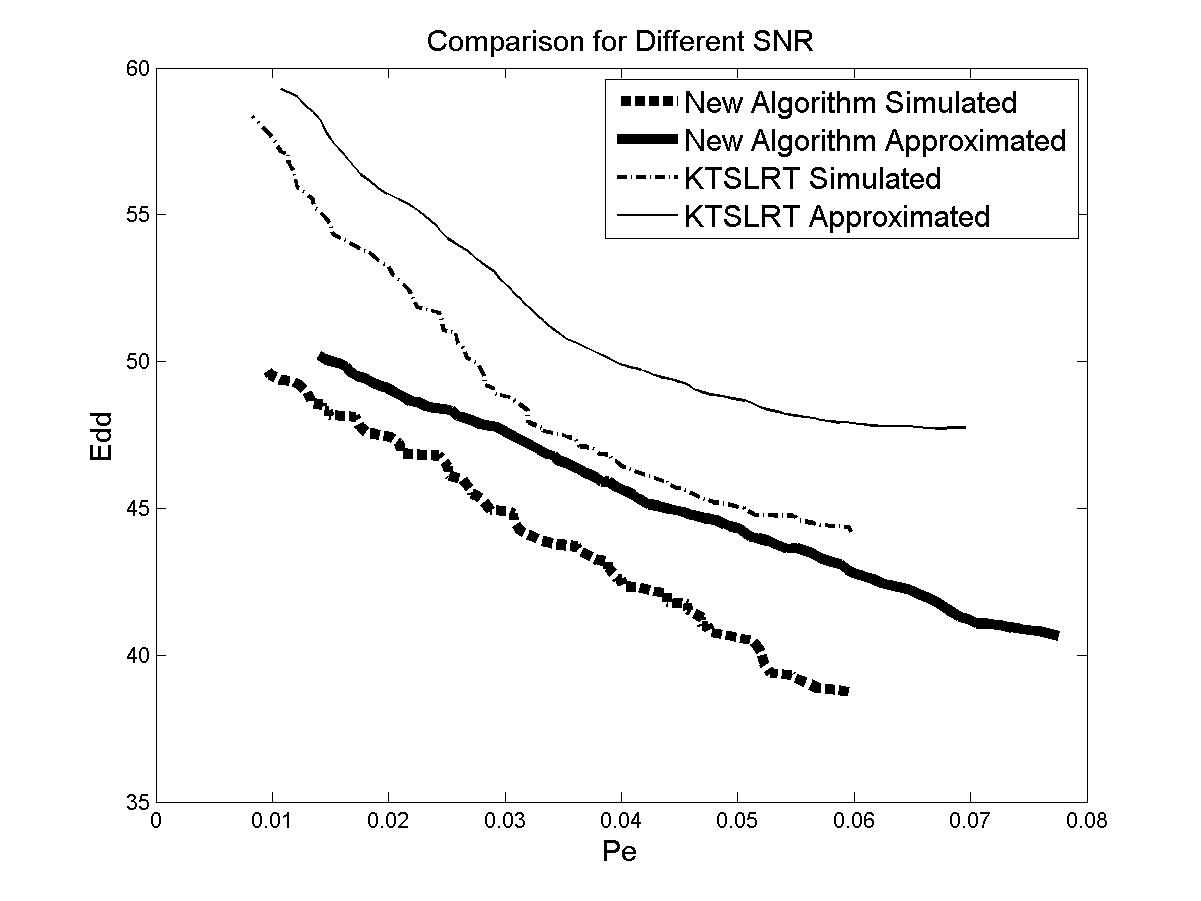}
                \label{fig:gull}
        \end{subfigure}
        \caption{Performance Comparison with KT-SLRT for Gaussian Distribution with different received SNRs}\label{fig:animals}
\end{figure}
\section{Further Generalizations}
Let us now consider a generalization of the problem, in which $P_{0}$ is not exactly known. Specifically, the hypothesis testing problem we now consider is:
\begin{equation}
H_{0} : P \in \{P_{0}' : D(P_{0}'||P_{0}) \leq \gamma\lambda\},\mbox{ for some }0\leq\gamma<1.
\end{equation}
\begin{equation*}
H_{1} : P \in \{P_{1}': D(P_{1}'||P_{0})\geq\lambda\mbox{ and } H(P_{1}')>H(P_{0}'),\end{equation*} for all $P_{0}'\in H_{0}\}$
\\The detection algorithm remains the same except that now we write the test statistic at the local node $l$ as
\begin{equation*}
\widetilde{W}_{k,l} = \widetilde{W}_{k-1,l}-\log\hat{P}_{0}(X_{k,l})-H(\hat{P}_{0})-\upsilon\lambda.
\end{equation*}  
For good performance we should pick $\hat{P}_{0}$ from the class in (5) and choose $\upsilon$ carefully. We elaborate on this in the following.
\par Let us try to justify this problem statement from a practical CR standpoint. In a CR setup, $H_{0}$ actually indicates the presence of only noise, while under $H_{1}$, the observatios are signal $+$ noise. Due to electromagnetic interference, the receiver noise can be changing with time (\cite{sahai}). Thus we assume that the noise power $P_{N}$ is  bounded as $\sigma_{N,L}^{2}\leq P_{N}\leq \sigma_{N,H}^{2}$. Similarly, let the signal power be bounded as $\sigma_{S,L}^{2}\leq P_{S}\leq \sigma_{S,H}^{2}$. Now we formulate these constraints in the form (5) where we should select appropriate $P_{0}$, $\lambda$ and $\gamma$. We will compute these assuming we are limiting ourselves to Gaussian distributions but will see that these work well in general.
\par We take, $P_{0}\sim\mathcal{N}(0,\sigma_{0}^{2})$, with $\sigma_{0}$ determined from the given bounds as follows.
\par Given two Gaussian distributions $Q_{0}$ and $Q_{1}$ with zero mean and variances $\sigma_{0}^{2}$ and $\sigma_{1}^{2}$ respectively,\\ \\
$D(Q_{1}||Q_{0}) = \displaystyle\ln\frac{\sigma_{0}}{\sigma_{1}}+\frac{1}{2}(\frac{\sigma_{1}^{2}}{\sigma_{0}^{2}}-1)$\\
Let $f(\sigma)\triangleq\displaystyle\ln\frac{\sigma_{0}}{\sigma}+\frac{1}{2}(\frac{\sigma^{2}}{\sigma_{0}^{2}}-1)$. We choose $\sigma_{0}$ such that $f(\sigma_{N,L})=f(\sigma_{N,H})$. This can be achieved for some $\sigma_{0}\in (\sigma_{N,L},\sigma_{N,H})$, since $f$ is convex with a minimum at $\sigma_{0}$. This choice ensures that $P_{0}$ is at some sort of a "centre" of the class of distributions under consideration in $H_{0}$. We now choose $\gamma\lambda\triangleq f(\sigma_{N,L})=f(\sigma_{N,H})$.
\par For the class of distributions considered under $H_{1}$, \begin{equation*}\sigma_{N,L}^{2}+\sigma_{S,L}^{2}\leq E[X^{2}]\leq \sigma_{N,H}^{2}+\sigma_{S,H}^{2}.\end{equation*} We take, \begin{equation*}\lambda\triangleq\displaystyle\inf_{\sigma^{2}\in (\sigma_{N,L}^{2}+\sigma_{S,L}^{2},\sigma_{N,H}^{2}+\sigma_{S,H}^{2})}f(\sigma)=f(\sqrt{\sigma_{N,L}^{2}+\sigma_{S,L}^{2}}).\end{equation*}
\par Next we compute $\hat{P}_{0}$. If the $X_{k,l}$ has distribution $P_{i}'$ for $i=0,1$, then the drift at the local nodes is $D(P_{0}'||\hat{P}_{0})+H(P_{0}')-H(\hat{P_{0}})-\upsilon\lambda$ under $H_{0}$, and $D(P_{1}'||\hat{P}_{0}) + H(P_{1}') - H(\hat{P_{0}})-\upsilon\lambda$ under $H_{1}$. This drift is an important parameter in determining the algorithm performance and will decide $\hat{P}_{0}$. 
\par Let $W_{i}$ be the cost of rejecting $H_{i}$ wrongly, and $c$ be the cost of taking each observation. Then, Bayes risk for the test is given (\cite{estimation}) by \\$\mathcal{R}_{c}(\delta)=\displaystyle \sum_{i=0}^{1}\pi^{i}[W_{i}P_{i}(\mbox{ reject }H_{i}) + cE_{i}(N)]$, where $\pi^{i}$ is the prior probability of $H_{i}$.\\
Taking the same thresholds as in Section V and using Theorems 5.1 and 5.2, \\ \\
$\displaystyle\lim_{c\to 0}\frac{\mathcal{R}_{c}(\delta)}{c|\log c|}$\\ \\ $\leq \frac{\pi^{0}}{L[-D(P_{0}'||\hat{P}_{0})-H(P_{0}')+H(\hat{P_{0}})+\upsilon\lambda]}(1-\frac{\theta_{0}}{\Delta(\mathcal{A}^{0})})+$\\$\frac{\pi^{1}}{L[D(P_{1}'||\hat{P}_{0}) + H(P_{1}') - H(\hat{P_{0}})-\upsilon\lambda]}(1+\frac{\theta_{1}}{\Delta(\mathcal{A}^{1})})$\begin{equation} - \frac{\pi^{0}}{\Delta(\mathcal{A}^{0})}+\frac{\pi^{1}}{\Delta(\mathcal{A}^{1})}.\end{equation} 
Following a minimax approach, we first maximize the above expression with respect to $P_{0}'$ and $P_{1}'$, and then minimize the resulting maximal risk w.r.t. $\hat{P}_{0}$ and $\upsilon$. As noted before, we achieve this optimization limiting ourselves to only Gaussian family. 
\par The second term in (6) is maximized when $D(P_{1}'||\hat{P}_{0}) + h(P_{1}')$ is minimized. Let us denote the variance of $\hat{P}_{0}$ by $\Gamma$. Now, the variances of all eligible $P_{1}'$s are greater than $\Gamma^{2}$. Hence, $D(P_{1}'||\hat{P}_{0}) + h(P_{1}')$ is minimized when $P_{1}'$ has the least possible variance, i.e. $\sigma_{N,L}^{2}+\sigma_{S,L}^{2}$. Using $\mathcal{N}(0,\sigma_{N,L}^{2}+\sigma_{S,L}^{2})$ in place of $P_{1}'$, the second term in (6) becomes (after simplification), \begin{equation*}\displaystyle\frac{(\pi^{1}/L)(1+\frac{\theta_{1}}{\Delta(\mathcal{A}^{1})})}{\frac{1}{2}(\frac{\sigma_{N,L}^{2}+\sigma_{S,L}^{2}}{\Gamma^{2}}-1)-\upsilon\lambda}.\end{equation*}
\par Similarly, to maximize the first term in (6), we have to minimize $D(P_{0}'||\hat{P}_{0}) + H(P_{0}')$ w.r.t. $P_{0}'$. After this, the first term becomes $\displaystyle\frac{(\pi^{0}/L)(1-\frac{\theta_{0}}{\Delta(\mathcal{A}^{0})})}{\upsilon\lambda-\frac{1}{2}(\frac{\sigma_{N,H}^{2}}{\Gamma^{2}}-1)}$.  \begin{equation*}\mbox{Taking  }x\triangleq\displaystyle\frac{1}{\Gamma^{2}}, y\triangleq\upsilon\lambda, a=\sigma_{N,H}^{2},b=\sigma_{N,L}^{2}+\sigma_{S,L}^{2},\end{equation*}\begin{equation} A= (\pi^{0}/L)(1-\frac{\theta_{0}}{\Delta(\mathcal{A}^{0})})\mbox{  and  }B=(\pi^{1}/L)(1+\frac{\theta_{1}}{\Delta(\mathcal{A}^{1})}),\end{equation}
the non-constant part of the optimized expression (6) can be written as a function of $x$ and $y$ in the form, \begin{equation*}g(x,y)=\displaystyle\frac{A}{y+\frac{1}{2}-\frac{1}{2}ax}+\frac{B}{\frac{1}{2}bx - y -\frac{1}{2}}.\end{equation*}
Minimizing this w.r.t. $y$ yields,\begin{equation}y_{opt} = \displaystyle\frac{1}{2}\frac{\sqrt{A}(bx-1)+\sqrt{B}(ax-1)}{\sqrt{A}+\sqrt{B}}\end{equation}
Together with this, we can choose $x\in\displaystyle (\frac{1}{\sigma_{N,H}^{2}},\frac{1}{\sigma_{N,L}^{2}})$.
\par In the following, we demonstrate the advantage of optimizing the above parmeters on the examples considered in Section VI. The bounds on the noise and signal power were chosen in each case such that the distributions specified in Section VI satisfy those constraints. Also, the thresholds were chosen the same as before.\par For the following simulations, we have taken \\$\Gamma^{2}=\displaystyle\frac{\sigma_{N,L}^{2}+\sigma_{N,H}^{2}}{2}$ and determined $y_{opt}$ in accordance with (8). \\ 
For Gaussian distribution, $P_{0}'\equiv\mathcal{N}(0,1)$, $P_{1}\equiv\mathcal{N}(0,5)$\\ 
For Lognormal distribution, $P_{0}'\equiv\log\mathcal{N}(0,3)$, $P_{1}\equiv\log\mathcal{N}(3,3)$\\ 
For Pareto distribution, $P_{0}'\equiv\mathcal{P}(10,2)$, $P_{1}\equiv\mathcal{P}(3,2)$\\ 
We compare the performances in Figs. 8-10. We see that the optimized version performs noticeably better, even for distributions other than Gaussian.
\begin{figure}[h]        
                \centering
                \includegraphics[width=.45\textwidth,height=4cm]{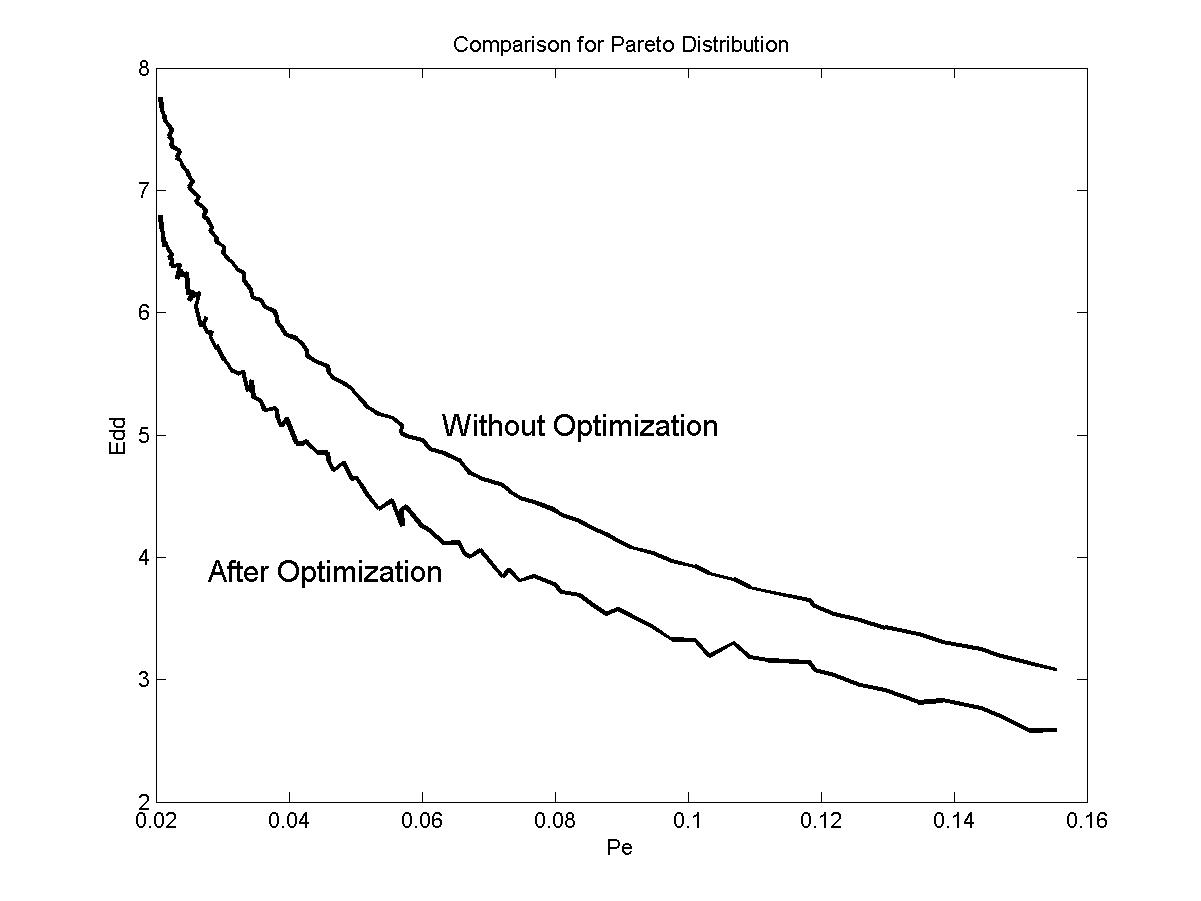}
                \caption{Optimization for Pareto Distribution}
                \label{fig:gull}
\end{figure}
\begin{figure}[h]        
                \vspace{-20pt}
                \centering
                \includegraphics[width=.45\textwidth,height=4cm]{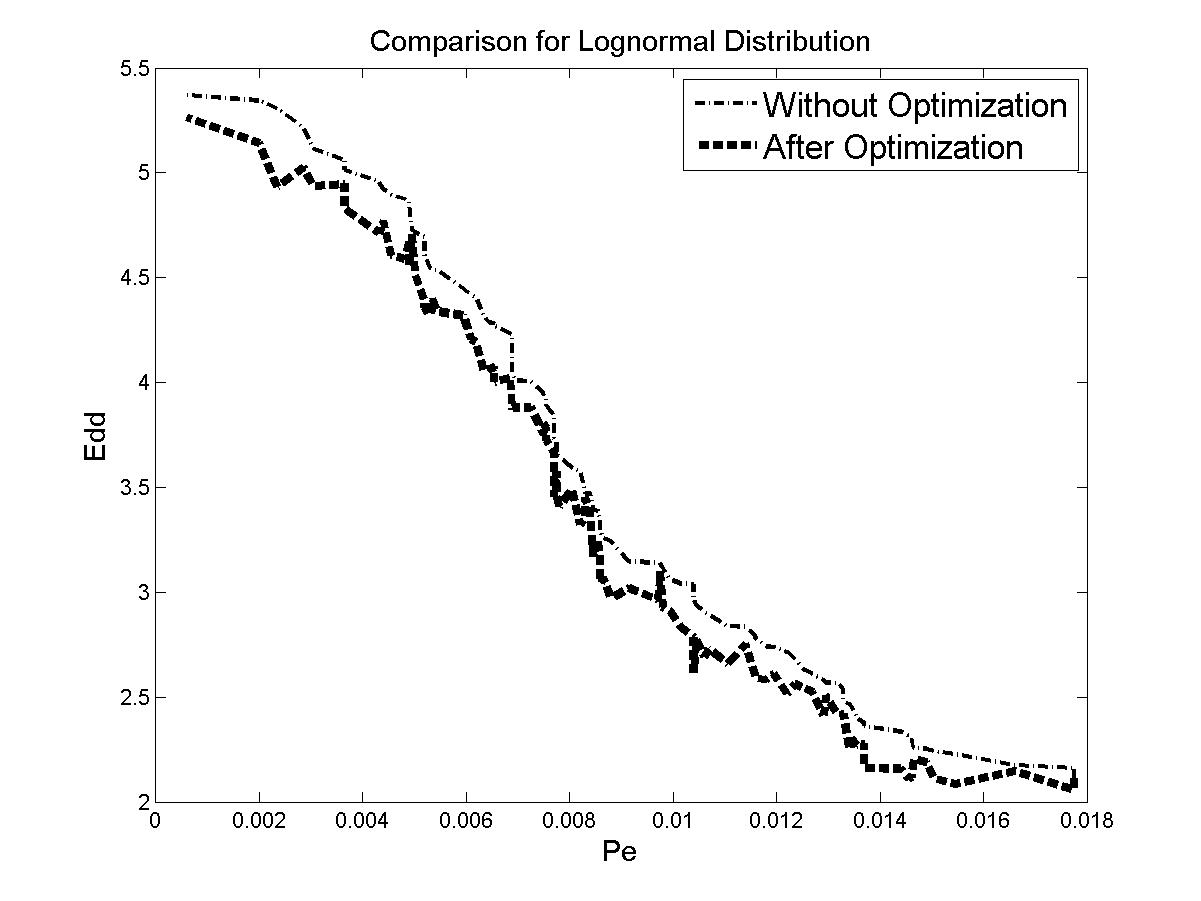}
                \caption{Optimization for Lognormal Distribution}
                \label{fig:gull}
\end{figure}
\begin{figure}[h]        
                \vspace{-15pt}
                \centering
                \includegraphics[width=.45\textwidth,height=4cm]{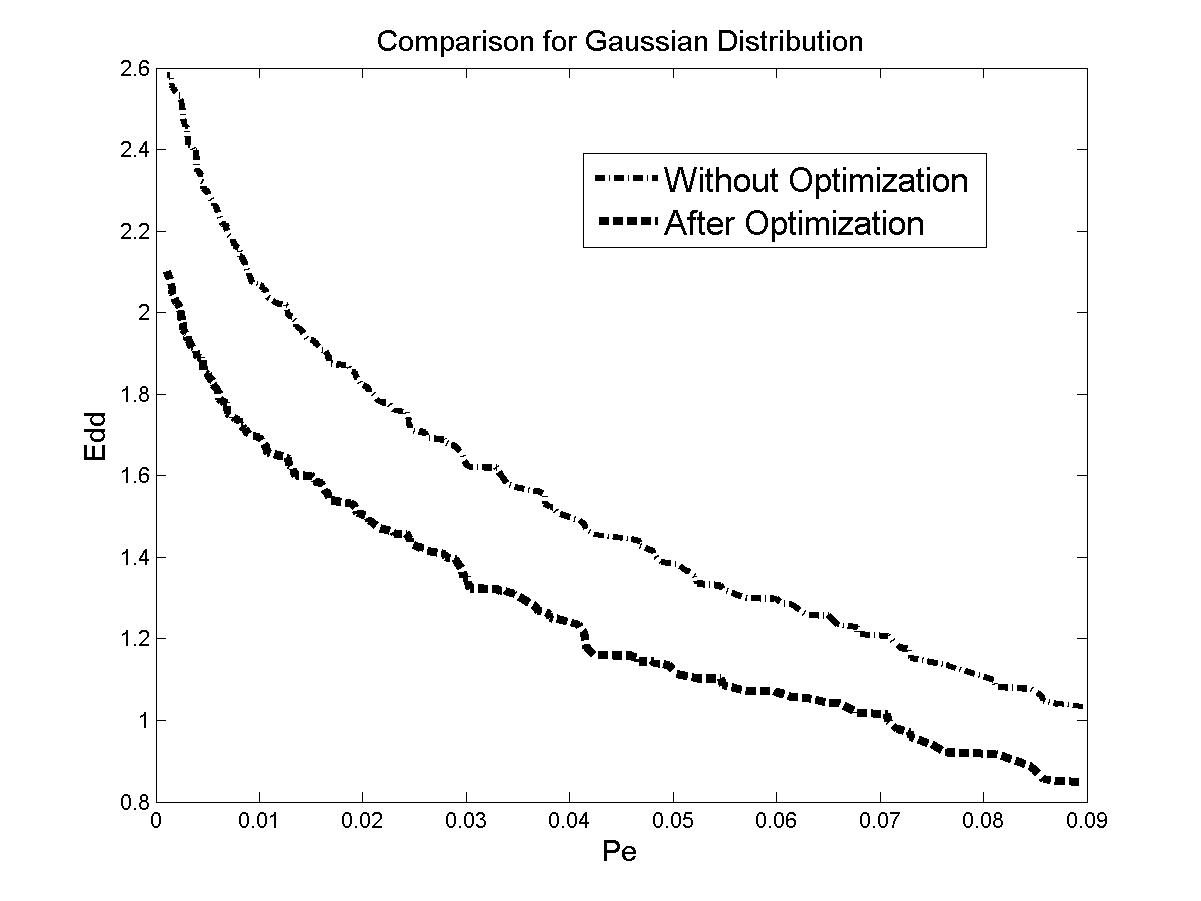}
                \caption{Optimization for Gaussian Distribution}
                \label{fig:gull}
\end{figure}
\section{Conclusions}
We have developed a new distributed sequential algorithm for detection, where under one of the hypotheses, the distribution can belong to a nonparametric family. This can be useful for spectrum sensing in Cognitive Radios. This algorithm is shown to perform better than a previous algorithm which was known to perform well and is also easier to implement. We have also obtained its performance approximately and studied asymptotic performance. The approximations match with the simulations better than the asymptotics. The asymptotics are comparable to SPRT and other known algorithms even though it is in the non-parametric setup.
\bibliographystyle{ieeetr}	
\bibliography{ref_mimo}		

\end{document}